\documentclass[%
 reprint,
 amsmath,amssymb,
 aps,
]{revtex4-1}

\usepackage{graphicx}
\usepackage{dcolumn}
\usepackage{bm}
\usepackage{float}

\newcommand{\micron}{\,\mathrm{\mu m}}
\newcommand{\nm}{\,\mathrm{nm}}
\newcommand{\cc}{\,\mathrm{cm^{-3}}}
\newcommand{\Wcm}{\,\mathrm{W/cm}^2}

\begin{document}


\title{High-charge relativistic electron bunches from a kHz laser-plasma accelerator}

\author{D. Gustas, D. Gu\'enot, A. Vernier, S. Dutt, F. B\"ohle,  R. Lopez-Martens, A. Lifschitz and J. Faure}
\affiliation{%
 LOA, ENSTA ParisTech, CNRS, Ecole polytechnique, Univ. Paris-Saclay, Palaiseau, France\\
}%

\date{\today}

\begin{abstract}
We report on electron wakefield acceleration in the resonant bubble regime with few-millijoule near-single-cycle laser pulses at a kilohertz repetition rate. Using very tight focusing of the laser pulse in conjunction with microscale supersonic gas jets, we demonstrate a stable relativistic electron source with a high charge per pulse up to $24$~pC/shot. The corresponding average current is 24~nA, making this kilohertz electron source useful for various applications.  
\end{abstract}

\pacs{Valid PACS appear here}
\maketitle



Laser wakefield acceleration (LWFA) is an established technique for producing high-energy electrons over minuscule distances \cite{esar09}. Due to its ability to generate ultrashort particle bunches \cite{lund11} as well as a wide range of secondary radiation with small source sizes \cite{rous04, knei10, fuch09, taph12}, the method is often considered for many applications in industry, material science, nuclear physics or medicine \cite{albe10}. However, most LWFA experiments are currently performed using 100 TW class laser systems at low repetition rate ($\leq 1$ Hz), which limits their practical use. Increasing the repetition rate is important for a wide range of reasons: (i) it permits reaching a higher level of stability; (ii) it enables rapid averaging over many shots, thereby significantly increasing the signal-to-noise ratio; (iii) it can boost the average current of the electron source by several orders of magnitude. Irradiation-based applications such as in medical treatment \cite{malk08} or electronics hardness studies \cite{hidd17} would directly benefit from a high average current because the required dose could be delivered in a much shorter time. Applications relying on a pump-probe scheme, such as ultrafast electron diffraction \cite{scia11, he16} or pulsed radiolysis \cite{muro08}, would greatly benefit from the higher stability and the improved signal-to-noise ratio.

To address these points, some of the recent work has been dedicated to developing high-repetition rate laser-plasma accelerators driven by low-energy laser pulses, in the 1-10~mJ range. Initial attempts produced sub-relativistic electrons with relatively low charge and relied on density down ramp injection \cite{he13, beau15}. The first MeV-scale accelerator at a kHz repetition rate was obtained using ultra-high density gas targets; it operated in the self-modulated regime, resulting in a fairly divergent beam with a maxwellian energy distribution \cite{salehi17}. To improve the source performance, our group has recently adapted the well-known "bubble" regime \cite{pukh02} configuration for few-millijoule laser pulses by compressing them nearly to a single optical cycle, or below 4 fs. Higher quality beams were obtained, with divergences of $\sim40$~mrad, stable peaked energy distribution at $\sim 5$~MeV and charges of hundreds of fC \cite{guenot17}. Simulations showed that electrons were injected via ionization \cite{pak10, mcgu10} of the K-shell electrons in nitrogen, yielding ultrashort relativistic bunches generated in the first arch of the wakefield \cite{lund13}. Despite kHz repetition rate, however, the electron source displayed relatively high charge fluctuations, indicating proximity to the injection threshold \cite{mang12}.

In the present article, we circumvent this problem by driving the accelerator at higher laser intensity. The interaction of the laser with the plasma medium is optimized by using innovative microscale supersonic gas jets providing higher density gradients and shorter plasma lengths. We demonstrate a laser-plasma accelerator running at kHz, producing few MeV electron beams with stable beam charge up to 24~pC/shot, \textit{i.e.} a two order of magnitude improvement compared to previous results. This yields an average current of 24~nA, the largest ever measured in a laser-plasma accelerator. In section \ref{sec:jets}, we discuss design issues and characterization of the microscale gas jets. In section \ref{sec:experiment}, we show the results of the experiment and discuss them in section \ref{sec:simu} on the basis of Particle In Cell (PIC) simulations.

\section{\label{sec:jets}Microscale supersonic jets}
\begin{figure}[t!]
\includegraphics[width=8cm]{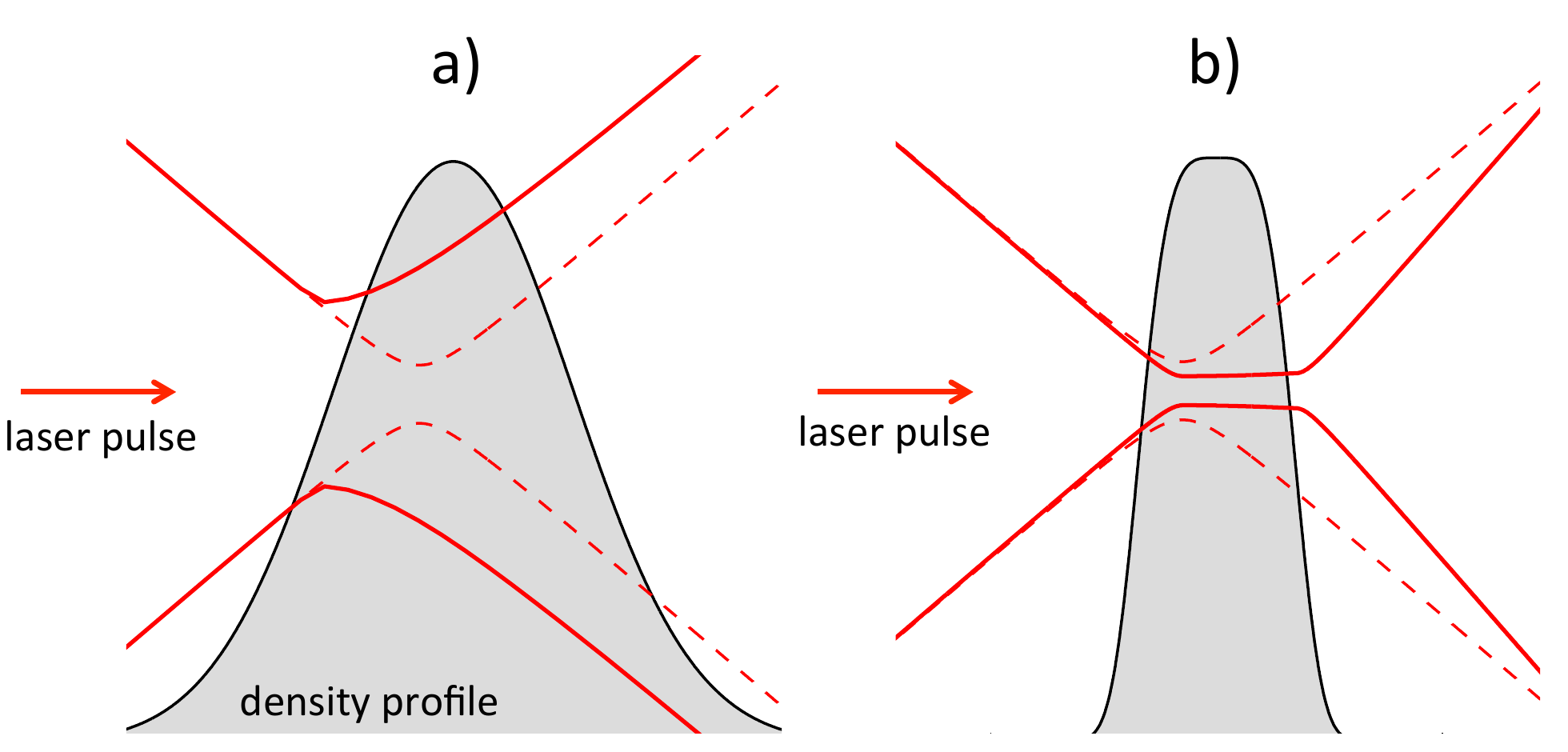}
\caption{Schematic of beam propagation issues in microscale jets. The dashed line represent the vacuum laser beam whereas the full line shows the beam size considering plasma effects. (a) The density gradients are large compared to $z_R$ preventing the laser beam from reaching high intensity in the jet. (b) With sharper density gradients, coupling into the jet is optimized and the laser beam can reach higher intensities through self-focusing. }
\label{schema}
\end{figure}

The laser-plasma accelerator is operated near the bubble regime which is known to produce small divergence beams with quasi-monoenergetic distributions \cite{mang04, gedd04, faur04}. This regime may be accessed once the light is focused to relativistic intensities $\,\approx 10^{18}-10^{19}\Wcm$, and the resonance condition is satisfied \cite{lu07}: $c\tau \approx w_0 \approx \lambda_p/2$, where $c$ is the speed of light, $\tau$ is the FWHM pulse duration, $w_0$ is the laser beam waist, and $\lambda_p$ is the target plasma wavelength. For few mJ laser systems, these conditions can be fulfilled provided that the laser pulse is extremely short, typically $<4$~fs, and is focused tightly, $w_0\approx2-3\micron$ in a high density plasma of $n_e\approx 1-2\times10^{20}\cc$ for a $800\nm$ laser wavelength. At this high density, the dephasing length and the pump depletion length are very short \cite{lu07}, of the order of $\approx20\micron$, which calls for the use of very thin gas targets.  In addition, while laser pulses with octave-spanning spectrum are used, strong dispersion effects similarly limit the high-intensity interaction to no more than few tens of microns \cite{guenot17, beau14}. Finally, the most important issue originates from laser beam propagation: for a waist of $w_0=2\micron$, the Rayleigh length is estimated to be $z_R=16\micron$. Therefore, sharp density gradients are crucial for optimizing the coupling of the laser pulses into the jet and avoiding ionization-induced defocusing \cite{rae93}. This effect is particularly detrimental when using high-Z gases, as in our experiment. Fig.~\ref{schema}a illustrates beam propagation in the case where the density gradient is longer than the Rayleigh length. In this arrangement, ionization-induced defocusing prevents the laser from reaching intensities required to drive a large amplitude wakefield. In the contrary case of Fig.~\ref{schema}b, the density gradient is short enough to allow the laser beam to self-focus in the jet, resulting in the excitation of a strong wakefield. These considerations clearly indicate that gas nozzles providing jets of $\approx100\micron$ with sharp density gradients are ideal.

There are also more practical considerations that need to be considered for nozzle design. Firstly, the tip of the nozzle cannot be brought closer than $100\, \mu$m to the laser focus without getting damaged by the laser itself. Secondly, the nozzle needs to provide high density in a continuous gas flow in order to enable operation at high repetition rates. This is considerably challenging for the vacuum pumping system as it needs to keep the background pressure in the chamber below $10^{-2}$ mbar. Consequently, the nozzle should be designed in order to minimize the mass flow while maximizing the density at heights above $100\micron$. This calls for microscale supersonic nozzles that are able to provide high densities well above the tip opening.

\begin{figure}[t!]
\includegraphics[width=7.5cm]{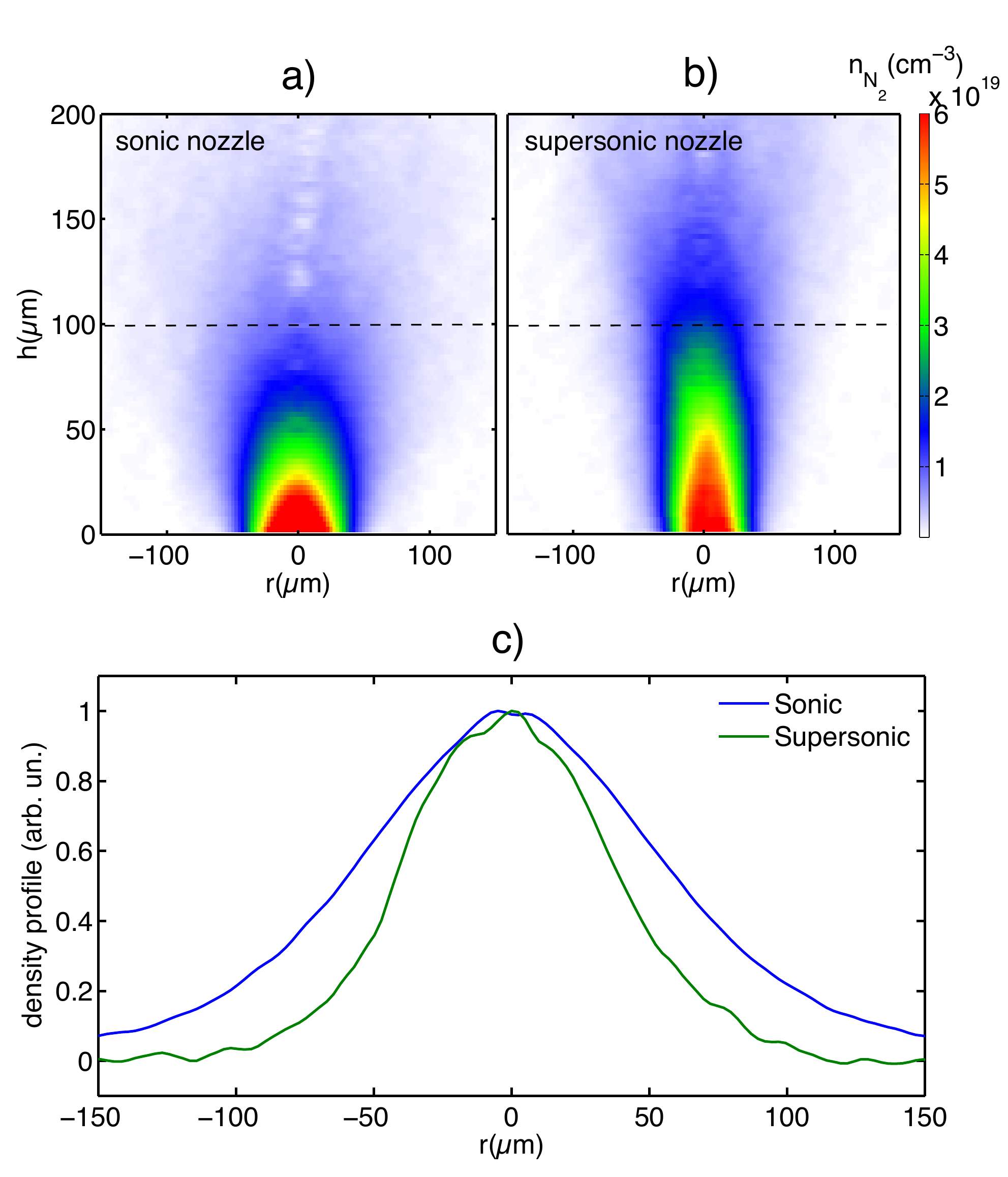}
\caption{Molecular density map $n_{N_2}(r,h)$ (a) for the subsonic nozzle with backing pressure $P=12$ bar and (b) for the supersonic nozzle, with backing pressure $P=60$ bar. Both cases lead to similar background pressure in the vacuum chamber but the peak density at $h=100\micron$ is $n_{N_2}=1.8\times10^{19}\cc$ for the supersonic jet and $n_{N_2}=8\times10^{18}\cc$ for the subsonic nozzle. (c) Normalized density profiles obtained at height $h=100\micron$ (dashed line in (a) and (b)). The $1/e$ width is $51\micron$ ($80\micron$) for the supersonic (subsonic) nozzle.}
\label{target}
\end{figure}

In Fig.~\ref{target}, we compare a $100\, \mu$m-diameter nozzle, providing a subsonic flow to a supersonic conical De Laval nozzle with a $40\, \mu$m throat \cite{schm12}, specially manufactured for this experiment by micro spark erosion. The $N_2$ gas jets are characterized with a quadriwave lateral shearing interferometer (SID4 HR by PHASICS). The density maps are obtained via Abel inversion of the measured phase maps. The backing pressure ($P=12$ bar for the subsonic nozzle and $P=60$ bar for the supersonic nozzle) is chosen such that the background pressure in the vacuum chamber is similar in both cases, enabling a direct comparison of the density profiles. Clearly, the supersonic jet provides higher density above $100\micron$, while preserving a thinner length and sharper gradients compared to the subsonic nozzles. The supersonic jets fulfill the stringent experimental requirements of a high-repetition rate laser-plasma accelerator.

\section{\label{sec:experiment}Experimental results}

The experiment at LOA was performed using the Salle Noire laser system delivering 3.9 fs pulses ($\approx 1.5$ optical cycle at $\lambda_{0} \approx 800$ nm) at 1 kHz with 2.5 mJ of energy on target \cite{boeh14}. A pair of motorized fused-silica wedges could be adjusted in the beam path to introduce some predominantly second-order chirp. An f/2 parabola was utilized to focus the light into a near-Gaussian spot with dimensions $2.9 \times 2.5\, \mu$m (FWHM), implying an approximate Rayleigh range of $20-25\, \mu$m and a maximum vacuum intensity $I_{vac}\approx5\times10^{18}$ W/cm$^2$, estimated using the real focal spot image. An independently calibrated CsI(Tl) phosphor screen was used to measure the charge and visualize the electron spot. A pinhole and a pair of circular permanent magnets could be inserted into the beam path to measure particle spectra. Compared to our previous experiments \cite{guenot17}, we now operate well above the injection threshold by focusing the laser tighter and thus increasing the peak intensity by a factor of two. While this leads to a shorter Rayleigh length, we used the supersonic nozzles with sharp gradients in order to optimize coupling of the laser pulse into the gas jet. Nitrogen gas was used because each nitrogen molecule releases 10 electrons assuming immediate ionization of nitrogen to N$^{5+}$. Therefore, the required high electron density can be achieved while keeping the background pressure in the vacuum chamber at a reasonable level. It also gives the opportunity of ionization injection from K-shell electrons. The density profile experienced by the driver pulse depends on the backing pressure and the nozzle orifice distance to the optical axis (Fig.~\ref{target}a-b), which can be adjusted mechanically. The plasma profile can be well approximated by a Gaussian function characterized by its peak density and its $1/e$ width, see Fig.~\ref{target}c. 

Experiments with the subsonic jet yielded no relativistic electrons. As discussed earlier (see Fig.~\ref{schema}), this was likely due to ionization-induced beam propagation issues. Using the supersonic jet, such detrimental effects were clearly suppressed as a multi-pC electron beam could be obtained easily. 

\begin{figure}[t!]
\includegraphics[width=8.6cm]{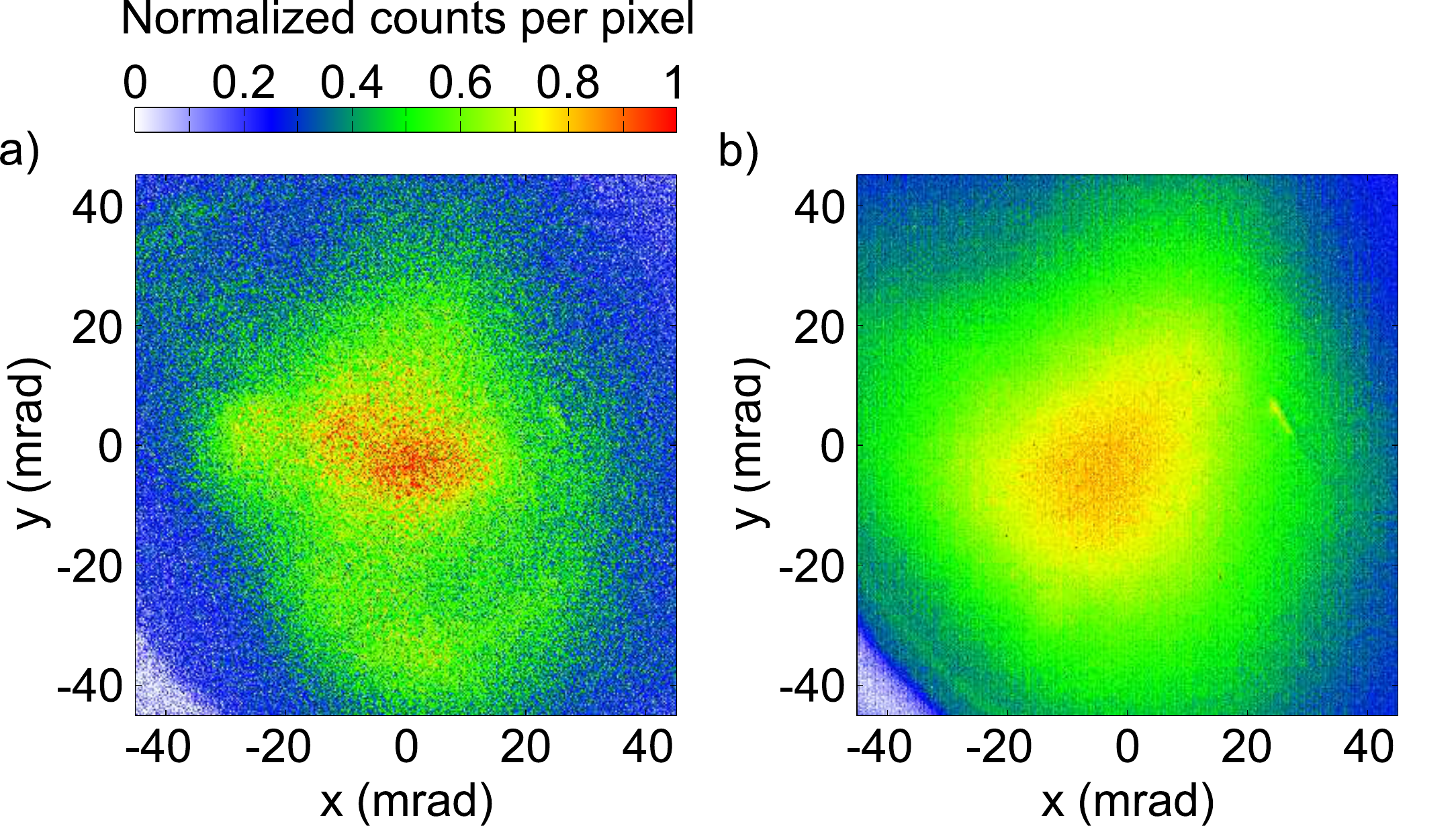}
\caption{Electron beam profiles corresponding to case 1. (a) Single-shot image. Divergence -- $44\times57$ mrad FWHM (b) 40-shot average, $74\times75$ mrad FWHM. Estimated charge -- 2.5 pC/shot ($\pm 14\%$ st. dev.).}
\label{beam}
\end{figure}

\begin{figure*}[t!]
\includegraphics[width=15.5cm]{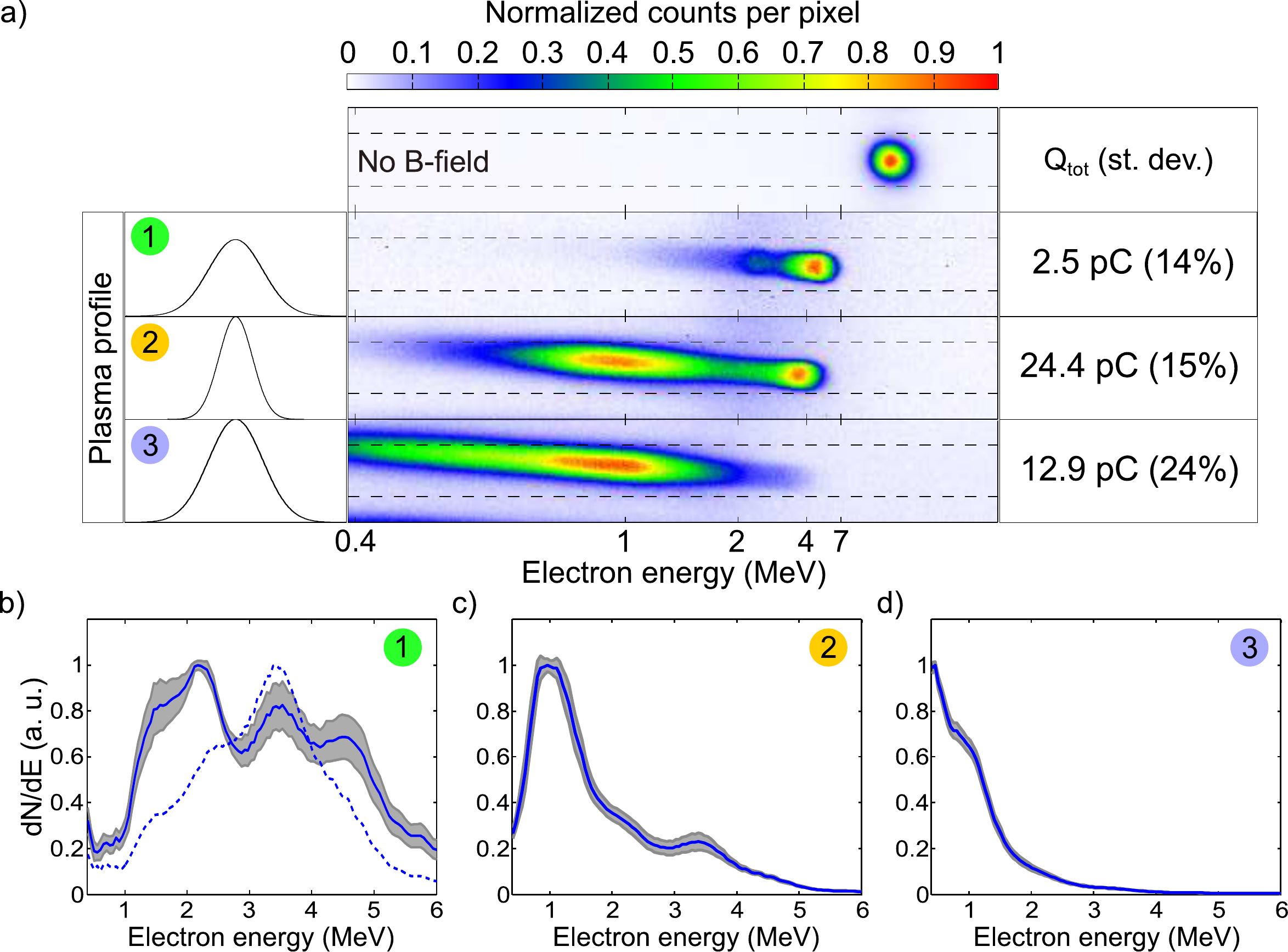}
\caption{Electron spectra and measured average charges for 3 different target density profiles from the supersonic jet. (a) Raw data and average charge values. Top row -- electron beam spatially filtered by a pinhole but not deviated by magnets. Case 1 -- spectrum obtained with peak electron density $1.45\times10^{20}$ cm$^{-3}$ and $1/e$ width of $65 \micron$. Case 2 corresponds to the case with $1.7\times10^{20}$ cm$^{-3}$ peak and $55\micron$ width, case 3 -- with $1.6\times10^{20}$ cm$^{-3}$ peak and $70\micron$ width. (b) Solid line -- deconvolved spectrum 1 with standard deviation (grey area). Dashed line -- deconvolved spectrum when the driver pulse was negatively chirped by $-4$~fs$^2$. (c) and (d) Deconvolved spectra 2 and 3 with corresponding standard deviations.}
\label{spectra}
\end{figure*}

We found the electron beam energy distribution to be very sensitive to the density profile. Fig.~\ref{spectra} shows the electron beam spectra obtained in three different cases corresponding to various densities and jet profiles. In case 1, a relativistic beam with a charge of 2.5 pC/shot was obtained by focusing the laser into the rising edge of a Gaussian plasma profile with peak electron density $1.45\times10^{20}$ cm$^{-3}$ and $1/e$ width of $65\micron$ (Fig. \ref{spectra}, case 1). The measured spectral distribution was nearly a plateau extending from 1.5 to 5 MeV (Fig. \ref{spectra}b, solid line). In this regime, we observed that the electron beam parameters could be varied by chirping the driver pulse. For example, introducing a slight negative chirp roughly preserved the total charge but produced a narrower energy spread, leaving only a peak at $\approx 3.5$ MeV ($-4$ fs$^2$, Fig. \ref{spectra}b, dashed line).

In case 2, a higher peak density ($ 1.7\times10^{20}$ cm$^{-3}$) and thinner profile ($1/e$ width of $ 55\micron$) was obtained by moving the nozzle closer to the optical axis. We observed an increase in charge by almost a whole order of magnitude to $\approx$ 24 pC, accompanied by the appearance of a very strong peak at around 1 MeV, while the previous high-energy spectral feature at 3-4 MeV was preserved (Fig. \ref{spectra}a, line 2). Small chirp variations did not introduce any obvious trends, suggesting the entire injection process was well above the threshold. This data shows that sharper gradients and higher densities are beneficial for optimizing the beam charge. Finally, in case 3, the density was similar as above ($1.6\times10^{20}$ cm$^{-3}$), but the {1/e} width was made larger than in case 2 ($70\micron$). A roughly twofold decrease in charge was recorded, together with the disappearance of the high energy feature and a lengthening of the low-energy tail (Fig. \ref{spectra}a, line 3, and Fig. \ref{spectra}d). 

This data demonstrates a large sensitivity to the plasma profile and peak density, that can in turn be used as tuning knobs to shape the electron energy distribution. With supersonic jets, the accelerator now operates in a stable mode: the energy distribution is rather steady (see the grey lines in Fig.~\ref{spectra}b-d indicating the standard deviation of the distribution) and the shot-to-shot charge RMS fluctuations are in the $15\%$ range. Typical electron beam profiles are shown in Fig.~\ref{beam}: sub-60 mrad FWHM divergence is obtained. The comparison between single shot images and averaged images indicate that the beam pointing fluctuations are only a fraction of the beam divergence.

\section{\label{sec:simu}PIC simulations and discussion}

To get an insight into what types of injection mechanisms might be taking part in the process, we performed Particle-in-Cell (PIC) simulations using Calder-Circ \cite{lifs09}, a fully electromagnetic 3D code based on cylindrical coordinates $(r,z)$ and Fourier decomposition in the poloidal direction. The simulations were performed using a mesh with $\Delta z=0.1\,k_0^{-1}$ and $\Delta r=0.5\,k_0^{-1}$ (where $k_0$ is the laser wave vector), and the first five Fourier modes. We started with pure neutral nitrogen, which then experienced tunnel ionization. The neutral N gas density profile was a Gaussian with a peak value of $1/5\times1.7\times10^{20}$ cm$^{-3}$ and a $1/e$ width of $55\micron$, corresponding the experimental case 2. The number of macro-particles per cell before ionization was set to 500, which corresponds to $500\times5=2500$ macro-electrons per cell in the region of full ionization of the L-shell of nitrogen. The temporal high-frequency laser field, its peak normalized amplitude ($a_0=1.44$) and the beam focal spot size ($2.7\, \mu$m FWHM) were also matched to experimental inputs. 

\begin{figure*}[ht!]
\includegraphics[width=15.5cm]{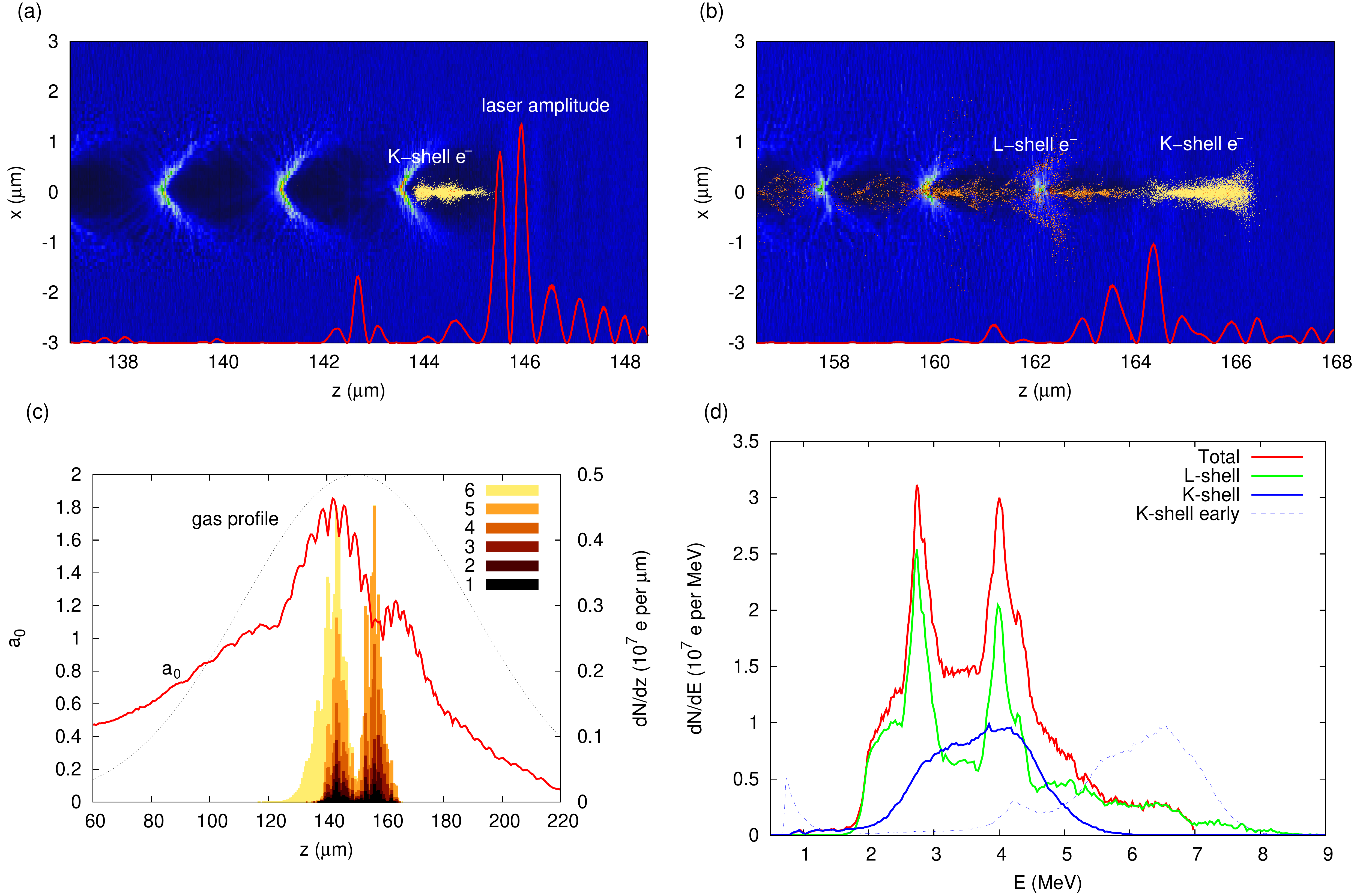}
\caption{ Results of 3D PIC simulations. (a)-(b) Electron density spatial distribution at two time steps. Trapped electrons are represented by dots. The red curve shows the laser E-field amplitude on the optical axis. (c) Red curve - evolution of the laser normalized amplitude. Histogram - particle trapping locations, decomposed into separate groups based on nitrogen ionization levels from which they originate. E.g. electrons marked by '1' originate from ionization  of N$^0$,  electrons '6' stem from ionization of N$^{5+}$, etc. (d) Final electron spectra of K-shell electrons ('6'), L-shell electrons ('1'-'5') and a sum of both. An early spectrum of K-shell electrons when the laser is around the middle of the gas jet is shown by the dashed line.}
\label{sims}
\end{figure*}

PIC simulations suggest that ionization injection is responsible for the trapping of electrons in the wakefield. This can be seen in Fig. \ref{sims}c, showing the histogram describing particle injection loci as well as evolution of the peak laser amplitude. The laser pulse self-focuses up to a maximum amplitude of $a_0=1.8$ around $10 \micron$ before the middle of the gas jet. It becomes intense enough to trigger ionization injection of electrons from the K-shell of nitrogen. A snapshot of the electron density spatial distribution at the end of the first injection is shown if Fig.~\ref{sims}a. As can be seen, these inner electrons (represented by yellow dots) are injected in the first wakefield period, making up a total charge of $\approx 5$ pC and extending over a $\approx2\micron$ distance. Right before the center of the gas jet ($z = 150\micron$), the laser intensity drops, stopping this first injection event.

The rapid evolution of the laser pulse then results in a second injection process. As described in \cite{beau14}, the laser pulse undergoes a strong redshift, causing its envelope to slip backwards because of the slower group velocity at red wavelengths. This results in a slow-down of the wakefield (see Fig.~\ref{sims}b) and triggers the second injection event. Indeed, the slower phase velocity of the wakefield enables efficient trapping of electrons even if the laser amplitude is significantly decreased  \cite{beau14}. The injection mechanism still relies on ionization even though the laser intensity is now too low to ionize K-shell electrons. We observe that the electrons originating from the $2p$ subshell are more often trapped than the ones coming from the $2s$ subshell (see Fig.~\ref{sims}c), confirming the role of ionization in this injection process. Contrary to the first injection event, a significant fraction of the electrons are trapped in the second bucket of the wakefield. The trapped charge due to this second injection event is close to the first one ($\approx 5$ pC). 

The final spectrum, shown in Fig. \ref{sims}d, extends from 2 MeV to 8 MeV, and possesses two peaks of roughly equal strength at 2.8 MeV and at 4 MeV. The total accelerated charge is 10 pC. Except the similar amplitudes of the two peaks, the simulation reproduces experimental results fairly closely. If only the electrons originating from the K-shell are considered, a broad peak between 2 and 5 MeV is obtained, similar to the dashed curve spectrum in Fig. \ref{spectra}b. We note that at the center of the gas jet, these electrons are faster, between 5 MeV and 8 MeV (dashed curve in Fig. \ref{sims}d). They are soon dephased because of the backward slip of the bubble and therefore start loosing energy. The peaks at 2.8 MeV and 4 MeV corresponds to L-shell particles injected into the second and first buckets respectively. The peak appears because of the rotation of the bunches in phase space ($z,p_z$) beyond the dephasing length. The RMS duration of the entire electron bunch is 10 fs, whereas its RMS divergence is 90 mrad.

The PIC simulation results suggest that two injection events occur in the experiment. At first, K-shell electrons are ionization-injected into the first bucket. Then, after massive self-focusing and reshaping of the driver pulse, the wakefield is slowed down, aiding the second injection process that results in the filling of several buckets. In the experimental case 1 (as labeled in Fig. \ref{spectra}), where the peak density is lower, this second injection might be mitigated, yielding lower charge, but higher energy and likely shorter electron bunches. At increased density, as in case 2, self-focusing is more pronounced, triggering the second injection event, which leads to significantly higher charge but lower energy electrons because of dephasing in the slower wakefield. If an overly wide N$_2$ gas jet is used, the K-shell electron peak might be lost, as in case 3, most probably due to stronger ionization-induced defocusing. We therefore suggest that precise control over the target profile might be a way to tune not only the injected charge or resultant spectrum, but also the number of buckets that are filled with accelerated electrons. To be fully validated, this hypothesis would still need to be tested experimentally through bunch length measurements \cite{lund11,lund13}.

\section{\label{sec:conclusion}Conclusion}

In summary, we produced a relativistic high-repetition rate electron wakefield accelerator with the highest average current ever recorded and with significantly enhanced stability. We have shown how its properties can be manipulated through small target width adjustments. The recorded broadband 24~pC source could be used to generate bright X-rays for radiography \cite{glin05} or to provide a laboratory-scale electronics damage testing environment for space industry, as the flux and spectrum are similar to those in Van Allen radiation belts \cite{hidd17}. In addition, these particle bunches could be of large interest for sub-10~fs jitter-free ultrafast electron diffraction experiments \cite{scia11, he16}. Previous work showed that with such large charge, the electron beam can be filtered spectrally and spatially to yield a sufficiently narrow energy spread and small emittance source for time-resolved matter studies \cite{faure16}. In conclusion, we believe the presented experiment plays an important role in the quest for providing stable, controllable and accessible particle sources to a wider user community.

\section*{Acknowledgements}
This work was funded by the European Research Council (ERC Starting Grant FEMTOELEC) under Contract No. 306708. Financial support from the R\'egion Ile-de-France (under contract SESAME-2012-ATTOLITE), the Agence Nationale pour la Recherche (under contracts ANR-11-EQPX-005-ATTOLAB and ANR-14-CE32-0011-03) and the Extreme Light Infrastructure-Hungary Non-Profit Ltd (under contract NLO3.6LOA) is gratefully acknowledged.


\begin{thebibliography}{32}%
\makeatletter
\providecommand \@ifxundefined [1]{%
 \@ifx{#1\undefined}
}%
\providecommand \@ifnum [1]{%
 \ifnum #1\expandafter \@firstoftwo
 \else \expandafter \@secondoftwo
 \fi
}%
\providecommand \@ifx [1]{%
 \ifx #1\expandafter \@firstoftwo
 \else \expandafter \@secondoftwo
 \fi
}%
\providecommand \natexlab [1]{#1}%
\providecommand \enquote  [1]{``#1''}%
\providecommand \bibnamefont  [1]{#1}%
\providecommand \bibfnamefont [1]{#1}%
\providecommand \citenamefont [1]{#1}%
\providecommand \href@noop [0]{\@secondoftwo}%
\providecommand \href [0]{\begingroup \@sanitize@url \@href}%
\providecommand \@href[1]{\@@startlink{#1}\@@href}%
\providecommand \@@href[1]{\endgroup#1\@@endlink}%
\providecommand \@sanitize@url [0]{\catcode `\\12\catcode `\$12\catcode
  `\&12\catcode `\#12\catcode `\^12\catcode `\_12\catcode `\%12\relax}%
\providecommand \@@startlink[1]{}%
\providecommand \@@endlink[0]{}%
\providecommand \url  [0]{\begingroup\@sanitize@url \@url }%
\providecommand \@url [1]{\endgroup\@href {#1}{\urlprefix }}%
\providecommand \urlprefix  [0]{URL }%
\providecommand \Eprint [0]{\href }%
\providecommand \doibase [0]{http://dx.doi.org/}%
\providecommand \selectlanguage [0]{\@gobble}%
\providecommand \bibinfo  [0]{\@secondoftwo}%
\providecommand \bibfield  [0]{\@secondoftwo}%
\providecommand \translation [1]{[#1]}%
\providecommand \BibitemOpen [0]{}%
\providecommand \bibitemStop [0]{}%
\providecommand \bibitemNoStop [0]{.\EOS\space}%
\providecommand \EOS [0]{\spacefactor3000\relax}%
\providecommand \BibitemShut  [1]{\csname bibitem#1\endcsname}%
\let\auto@bib@innerbib\@empty
\bibitem [{\citenamefont {Esarey}\ \emph {et~al.}(2009)\citenamefont {Esarey},
  \citenamefont {Schroeder},\ and\ \citenamefont {Leemans}}]{esar09}%
  \BibitemOpen
  \bibfield  {author} {\bibinfo {author} {\bibfnamefont {E.}~\bibnamefont
  {Esarey}}, \bibinfo {author} {\bibfnamefont {C.~B.}\ \bibnamefont
  {Schroeder}}, \ and\ \bibinfo {author} {\bibfnamefont {W.~P.}\ \bibnamefont
  {Leemans}},\ }\href@noop {} {\bibfield  {journal} {\bibinfo  {journal} {Rev.
  Mod. Phys.}\ }\textbf {\bibinfo {volume} {81}},\ \bibinfo {pages} {1229}
  (\bibinfo {year} {2009})}\BibitemShut {NoStop}%
\bibitem [{\citenamefont {Lundh}\ \emph {et~al.}(2011)\citenamefont {Lundh},
  \citenamefont {Lim}, \citenamefont {Rechatin}, \citenamefont {Ammoura},
  \citenamefont {Ben-Ismail}, \citenamefont {Davoine}, \citenamefont {Gallot},
  \citenamefont {Goddet}, \citenamefont {Lefebvre}, \citenamefont {Malka},\
  and\ \citenamefont {Faure}}]{lund11}%
  \BibitemOpen
  \bibfield  {author} {\bibinfo {author} {\bibfnamefont {O.}~\bibnamefont
  {Lundh}}, \bibinfo {author} {\bibfnamefont {J.}~\bibnamefont {Lim}}, \bibinfo
  {author} {\bibfnamefont {C.}~\bibnamefont {Rechatin}}, \bibinfo {author}
  {\bibfnamefont {L.}~\bibnamefont {Ammoura}}, \bibinfo {author} {\bibfnamefont
  {A.}~\bibnamefont {Ben-Ismail}}, \bibinfo {author} {\bibfnamefont
  {X.}~\bibnamefont {Davoine}}, \bibinfo {author} {\bibfnamefont
  {G.}~\bibnamefont {Gallot}}, \bibinfo {author} {\bibfnamefont {J.-P.}\
  \bibnamefont {Goddet}}, \bibinfo {author} {\bibfnamefont {E.}~\bibnamefont
  {Lefebvre}}, \bibinfo {author} {\bibfnamefont {V.}~\bibnamefont {Malka}}, \
  and\ \bibinfo {author} {\bibfnamefont {J.}~\bibnamefont {Faure}},\
  }\href@noop {} {\bibfield  {journal} {\bibinfo  {journal} {Nat. Phys.}\
  }\textbf {\bibinfo {volume} {7}},\ \bibinfo {pages} {219} (\bibinfo {year}
  {2011})}\BibitemShut {NoStop}%
\bibitem [{\citenamefont {Rousse}\ \emph {et~al.}(2004)\citenamefont {Rousse},
  \citenamefont {{Ta Phuoc}}, \citenamefont {Shah}, \citenamefont {Pukhov},
  \citenamefont {Lefebvre}, \citenamefont {Malka}, \citenamefont {Kiselev},
  \citenamefont {Burgy}, \citenamefont {Rousseau}, \citenamefont {Umstadter},\
  and\ \citenamefont {Hulin}}]{rous04}%
  \BibitemOpen
  \bibfield  {author} {\bibinfo {author} {\bibfnamefont {A.}~\bibnamefont
  {Rousse}}, \bibinfo {author} {\bibfnamefont {K.}~\bibnamefont {{Ta Phuoc}}},
  \bibinfo {author} {\bibfnamefont {R.}~\bibnamefont {Shah}}, \bibinfo {author}
  {\bibfnamefont {A.}~\bibnamefont {Pukhov}}, \bibinfo {author} {\bibfnamefont
  {E.}~\bibnamefont {Lefebvre}}, \bibinfo {author} {\bibfnamefont
  {V.}~\bibnamefont {Malka}}, \bibinfo {author} {\bibfnamefont
  {S.}~\bibnamefont {Kiselev}}, \bibinfo {author} {\bibfnamefont
  {F.}~\bibnamefont {Burgy}}, \bibinfo {author} {\bibfnamefont {J.-P.}\
  \bibnamefont {Rousseau}}, \bibinfo {author} {\bibfnamefont {D.}~\bibnamefont
  {Umstadter}}, \ and\ \bibinfo {author} {\bibfnamefont {D.}~\bibnamefont
  {Hulin}},\ }\href@noop {} {\bibfield  {journal} {\bibinfo  {journal} {Phys.
  Rev. Lett.}\ }\textbf {\bibinfo {volume} {93}},\ \bibinfo {pages} {135005}
  (\bibinfo {year} {2004})}\BibitemShut {NoStop}%
\bibitem [{\citenamefont {Kneip}\ \emph {et~al.}(2010)\citenamefont {Kneip},
  \citenamefont {McGuffey}, \citenamefont {Martins}, \citenamefont {Martins},
  \citenamefont {Bellei}, \citenamefont {Chvykov}, \citenamefont {Dollar},
  \citenamefont {Fonseca}, \citenamefont {Huntington}, \citenamefont
  {Kalintchenko}, \citenamefont {Maksimchuk}, \citenamefont {Mangles},
  \citenamefont {Matsuoka}, \citenamefont {Nagel}, \citenamefont {Palmer},
  \citenamefont {Schreiber}, \citenamefont {{Ta Phuoc}}, \citenamefont
  {Thomas}, \citenamefont {Yanovsky}, \citenamefont {Silva}, \citenamefont
  {Krushelnick},\ and\ \citenamefont {Najmudin}}]{knei10}%
  \BibitemOpen
  \bibfield  {author} {\bibinfo {author} {\bibfnamefont {S.}~\bibnamefont
  {Kneip}}, \bibinfo {author} {\bibfnamefont {C.}~\bibnamefont {McGuffey}},
  \bibinfo {author} {\bibfnamefont {J.~L.}\ \bibnamefont {Martins}}, \bibinfo
  {author} {\bibfnamefont {S.~F.}\ \bibnamefont {Martins}}, \bibinfo {author}
  {\bibfnamefont {C.}~\bibnamefont {Bellei}}, \bibinfo {author} {\bibfnamefont
  {V.}~\bibnamefont {Chvykov}}, \bibinfo {author} {\bibfnamefont
  {F.}~\bibnamefont {Dollar}}, \bibinfo {author} {\bibfnamefont
  {R.}~\bibnamefont {Fonseca}}, \bibinfo {author} {\bibfnamefont
  {C.}~\bibnamefont {Huntington}}, \bibinfo {author} {\bibfnamefont
  {G.}~\bibnamefont {Kalintchenko}}, \bibinfo {author} {\bibfnamefont
  {A.}~\bibnamefont {Maksimchuk}}, \bibinfo {author} {\bibfnamefont {S.~P.~D.}\
  \bibnamefont {Mangles}}, \bibinfo {author} {\bibfnamefont {T.}~\bibnamefont
  {Matsuoka}}, \bibinfo {author} {\bibfnamefont {S.~R.}\ \bibnamefont {Nagel}},
  \bibinfo {author} {\bibfnamefont {C.~A.~J.}\ \bibnamefont {Palmer}}, \bibinfo
  {author} {\bibfnamefont {J.}~\bibnamefont {Schreiber}}, \bibinfo {author}
  {\bibfnamefont {K.}~\bibnamefont {{Ta Phuoc}}}, \bibinfo {author}
  {\bibfnamefont {A.~G.~R.}\ \bibnamefont {Thomas}}, \bibinfo {author}
  {\bibfnamefont {V.}~\bibnamefont {Yanovsky}}, \bibinfo {author}
  {\bibfnamefont {L.~O.}\ \bibnamefont {Silva}}, \bibinfo {author}
  {\bibfnamefont {K.}~\bibnamefont {Krushelnick}}, \ and\ \bibinfo {author}
  {\bibfnamefont {Z.}~\bibnamefont {Najmudin}},\ }\href@noop {} {\bibfield
  {journal} {\bibinfo  {journal} {Nat. Phys.}\ }\textbf {\bibinfo {volume}
  {6}},\ \bibinfo {pages} {980} (\bibinfo {year} {2010})}\BibitemShut {NoStop}%
\bibitem [{\citenamefont {Fuchs}\ \emph {et~al.}(2009)\citenamefont {Fuchs},
  \citenamefont {Weingartner}, \citenamefont {Popp}, \citenamefont {Major},
  \citenamefont {Becker}, \citenamefont {Osterhoff}, \citenamefont {Cortrie},
  \citenamefont {Zeitler}, \citenamefont {H{\"o}rlein}, \citenamefont
  {Tsakiris}, \citenamefont {Schramm}, \citenamefont {Rowlands-Rees},
  \citenamefont {Hooker}, \citenamefont {Habs}, \citenamefont {Krausz},
  \citenamefont {Karsch},\ and\ \citenamefont {Gr{\"u}ner}}]{fuch09}%
  \BibitemOpen
  \bibfield  {author} {\bibinfo {author} {\bibfnamefont {M.}~\bibnamefont
  {Fuchs}}, \bibinfo {author} {\bibfnamefont {R.}~\bibnamefont {Weingartner}},
  \bibinfo {author} {\bibfnamefont {A.}~\bibnamefont {Popp}}, \bibinfo {author}
  {\bibfnamefont {Z.}~\bibnamefont {Major}}, \bibinfo {author} {\bibfnamefont
  {S.}~\bibnamefont {Becker}}, \bibinfo {author} {\bibfnamefont
  {J.}~\bibnamefont {Osterhoff}}, \bibinfo {author} {\bibfnamefont
  {I.}~\bibnamefont {Cortrie}}, \bibinfo {author} {\bibfnamefont
  {B.}~\bibnamefont {Zeitler}}, \bibinfo {author} {\bibfnamefont
  {R.}~\bibnamefont {H{\"o}rlein}}, \bibinfo {author} {\bibfnamefont {G.~D.}\
  \bibnamefont {Tsakiris}}, \bibinfo {author} {\bibfnamefont {U.}~\bibnamefont
  {Schramm}}, \bibinfo {author} {\bibfnamefont {T.~P.}\ \bibnamefont
  {Rowlands-Rees}}, \bibinfo {author} {\bibfnamefont {S.~M.}\ \bibnamefont
  {Hooker}}, \bibinfo {author} {\bibfnamefont {D.}~\bibnamefont {Habs}},
  \bibinfo {author} {\bibfnamefont {F.}~\bibnamefont {Krausz}}, \bibinfo
  {author} {\bibfnamefont {S.}~\bibnamefont {Karsch}}, \ and\ \bibinfo {author}
  {\bibfnamefont {F.}~\bibnamefont {Gr{\"u}ner}},\ }\href@noop {} {\bibfield
  {journal} {\bibinfo  {journal} {Nat. Phys.}\ }\textbf {\bibinfo {volume}
  {5}},\ \bibinfo {pages} {826} (\bibinfo {year} {2009})}\BibitemShut {NoStop}%
\bibitem [{\citenamefont {{Ta Phuoc}}\ \emph {et~al.}(2012)\citenamefont {{Ta
  Phuoc}}, \citenamefont {Corde}, \citenamefont {Thaury}, \citenamefont
  {Malka}, \citenamefont {Tafzi}, \citenamefont {Goddet}, \citenamefont
  {C.Shah}, \citenamefont {Sebban},\ and\ \citenamefont {Rousse}}]{taph12}%
  \BibitemOpen
  \bibfield  {author} {\bibinfo {author} {\bibfnamefont {K.}~\bibnamefont {{Ta
  Phuoc}}}, \bibinfo {author} {\bibfnamefont {S.}~\bibnamefont {Corde}},
  \bibinfo {author} {\bibfnamefont {C.}~\bibnamefont {Thaury}}, \bibinfo
  {author} {\bibfnamefont {V.}~\bibnamefont {Malka}}, \bibinfo {author}
  {\bibfnamefont {A.}~\bibnamefont {Tafzi}}, \bibinfo {author} {\bibfnamefont
  {J.-P.}\ \bibnamefont {Goddet}}, \bibinfo {author} {\bibfnamefont
  {R.}~\bibnamefont {C.Shah}}, \bibinfo {author} {\bibfnamefont
  {S.}~\bibnamefont {Sebban}}, \ and\ \bibinfo {author} {\bibfnamefont
  {A.}~\bibnamefont {Rousse}},\ }\href@noop {} {\bibfield  {journal} {\bibinfo
  {journal} {Nat. Photon.}\ }\textbf {\bibinfo {volume} {6}},\ \bibinfo {pages}
  {308} (\bibinfo {year} {2012})}\BibitemShut {NoStop}%
\bibitem [{\citenamefont {Albert}\ and\ \citenamefont {Thomas}(2016)}]{albe10}%
  \BibitemOpen
  \bibfield  {author} {\bibinfo {author} {\bibfnamefont {F.}~\bibnamefont
  {Albert}}\ and\ \bibinfo {author} {\bibfnamefont {A.~G.~R.}\ \bibnamefont
  {Thomas}},\ }\href@noop {} {\bibfield  {journal} {\bibinfo  {journal}
  {{Plasma Phys. Control. Fusion},}\ }\textbf {\bibinfo {volume} {58}},\
  \bibinfo {pages} {103001} (\bibinfo {year} {2016})}\BibitemShut {NoStop}%
\bibitem [{\citenamefont {Malka}\ \emph {et~al.}(2008)\citenamefont {Malka},
  \citenamefont {Faure}, \citenamefont {Gauduel}, \citenamefont {Lefebvre},
  \citenamefont {Rousse},\ and\ \citenamefont {{Ta Phuoc}}}]{malk08}%
  \BibitemOpen
  \bibfield  {author} {\bibinfo {author} {\bibfnamefont {V.}~\bibnamefont
  {Malka}}, \bibinfo {author} {\bibfnamefont {J.}~\bibnamefont {Faure}},
  \bibinfo {author} {\bibfnamefont {Y.~A.}\ \bibnamefont {Gauduel}}, \bibinfo
  {author} {\bibfnamefont {E.}~\bibnamefont {Lefebvre}}, \bibinfo {author}
  {\bibfnamefont {A.}~\bibnamefont {Rousse}}, \ and\ \bibinfo {author}
  {\bibfnamefont {K.}~\bibnamefont {{Ta Phuoc}}},\ }\href@noop {} {\bibfield
  {journal} {\bibinfo  {journal} {Nature Physics}\ }\textbf {\bibinfo {volume}
  {44}},\ \bibinfo {pages} {447} (\bibinfo {year} {2008})}\BibitemShut
  {NoStop}%
\bibitem [{\citenamefont {Hidding}\ \emph {et~al.}(2017)\citenamefont
  {Hidding}, \citenamefont {Karger}, \citenamefont {K{\"o}nigstein},
  \citenamefont {Pretzler}, \citenamefont {Manahan}, \citenamefont {McKenna},
  \citenamefont {Gray}, \citenamefont {Wilson}, \citenamefont {Wiggins},
  \citenamefont {Welsh}, \citenamefont {Beaton}, \citenamefont {Delinikolas},
  \citenamefont {Jaroszynski}, \citenamefont {Rosenzweig}, \citenamefont
  {Karmakar}, \citenamefont {Ferlet-Cavrois}, \citenamefont {Costantino},
  \citenamefont {Muschitiello},\ and\ \citenamefont {Daly}}]{hidd17}%
  \BibitemOpen
  \bibfield  {author} {\bibinfo {author} {\bibfnamefont {B.}~\bibnamefont
  {Hidding}}, \bibinfo {author} {\bibfnamefont {O.}~\bibnamefont {Karger}},
  \bibinfo {author} {\bibfnamefont {T.}~\bibnamefont {K{\"o}nigstein}},
  \bibinfo {author} {\bibfnamefont {G.}~\bibnamefont {Pretzler}}, \bibinfo
  {author} {\bibfnamefont {G.~G.}\ \bibnamefont {Manahan}}, \bibinfo {author}
  {\bibfnamefont {P.}~\bibnamefont {McKenna}}, \bibinfo {author} {\bibfnamefont
  {R.}~\bibnamefont {Gray}}, \bibinfo {author} {\bibfnamefont {R.}~\bibnamefont
  {Wilson}}, \bibinfo {author} {\bibfnamefont {S.~M.}\ \bibnamefont {Wiggins}},
  \bibinfo {author} {\bibfnamefont {G.~H.}\ \bibnamefont {Welsh}}, \bibinfo
  {author} {\bibfnamefont {A.}~\bibnamefont {Beaton}}, \bibinfo {author}
  {\bibfnamefont {P.}~\bibnamefont {Delinikolas}}, \bibinfo {author}
  {\bibfnamefont {D.~A.}\ \bibnamefont {Jaroszynski}}, \bibinfo {author}
  {\bibfnamefont {J.~B.}\ \bibnamefont {Rosenzweig}}, \bibinfo {author}
  {\bibfnamefont {A.}~\bibnamefont {Karmakar}}, \bibinfo {author}
  {\bibfnamefont {V.}~\bibnamefont {Ferlet-Cavrois}}, \bibinfo {author}
  {\bibfnamefont {A.}~\bibnamefont {Costantino}}, \bibinfo {author}
  {\bibfnamefont {M.}~\bibnamefont {Muschitiello}}, \ and\ \bibinfo {author}
  {\bibfnamefont {E.}~\bibnamefont {Daly}},\ }\href@noop {} {\bibfield
  {journal} {\bibinfo  {journal} {Sci. Rep.}\ }\textbf {\bibinfo {volume}
  {7}},\ \bibinfo {pages} {24354} (\bibinfo {year} {2017})}\BibitemShut
  {NoStop}%
\bibitem [{\citenamefont {Sciaini}\ and\ \citenamefont
  {Miller}(2011)}]{scia11}%
  \BibitemOpen
  \bibfield  {author} {\bibinfo {author} {\bibfnamefont {G.}~\bibnamefont
  {Sciaini}}\ and\ \bibinfo {author} {\bibfnamefont {R.~J.~D.}\ \bibnamefont
  {Miller}},\ }\href@noop {} {\bibfield  {journal} {\bibinfo  {journal} {Rep.
  Prog. Phys.}\ }\textbf {\bibinfo {volume} {74}},\ \bibinfo {pages} {096101}
  (\bibinfo {year} {2011})}\BibitemShut {NoStop}%
\bibitem [{\citenamefont {He}\ \emph {et~al.}(2016)\citenamefont {He},
  \citenamefont {Beaurepaire}, \citenamefont {Nees}, \citenamefont {Gall\'e},
  \citenamefont {Scott}, \citenamefont {P\'erez}, \citenamefont {Lagally},
  \citenamefont {Krushelnick}, \citenamefont {Thomas},\ and\ \citenamefont
  {Faure}}]{he16}%
  \BibitemOpen
  \bibfield  {author} {\bibinfo {author} {\bibfnamefont {Z.-H.}\ \bibnamefont
  {He}}, \bibinfo {author} {\bibfnamefont {B.}~\bibnamefont {Beaurepaire}},
  \bibinfo {author} {\bibfnamefont {J.~A.}\ \bibnamefont {Nees}}, \bibinfo
  {author} {\bibfnamefont {G.}~\bibnamefont {Gall\'e}}, \bibinfo {author}
  {\bibfnamefont {S.~A.}\ \bibnamefont {Scott}}, \bibinfo {author}
  {\bibfnamefont {J.~R.~S.}\ \bibnamefont {P\'erez}}, \bibinfo {author}
  {\bibfnamefont {M.~G.}\ \bibnamefont {Lagally}}, \bibinfo {author}
  {\bibfnamefont {K.}~\bibnamefont {Krushelnick}}, \bibinfo {author}
  {\bibfnamefont {A.~G.~R.}\ \bibnamefont {Thomas}}, \ and\ \bibinfo {author}
  {\bibfnamefont {J.}~\bibnamefont {Faure}},\ }\href@noop {} {\bibfield
  {journal} {\bibinfo  {journal} {Sci. Rep.}\ }\textbf {\bibinfo {volume}
  {6}},\ \bibinfo {pages} {36224} (\bibinfo {year} {2016})}\BibitemShut
  {NoStop}%
\bibitem [{\citenamefont {Muroya}\ \emph {et~al.}(2008)\citenamefont {Muroya},
  \citenamefont {Lin}, \citenamefont {Han}, \citenamefont {Kumagai},
  \citenamefont {Sakumi}, \citenamefont {Ueda},\ and\ \citenamefont
  {Katsumura}}]{muro08}%
  \BibitemOpen
  \bibfield  {author} {\bibinfo {author} {\bibfnamefont {Y.}~\bibnamefont
  {Muroya}}, \bibinfo {author} {\bibfnamefont {M.}~\bibnamefont {Lin}},
  \bibinfo {author} {\bibfnamefont {Z.}~\bibnamefont {Han}}, \bibinfo {author}
  {\bibfnamefont {Y.}~\bibnamefont {Kumagai}}, \bibinfo {author} {\bibfnamefont
  {A.}~\bibnamefont {Sakumi}}, \bibinfo {author} {\bibfnamefont
  {T.}~\bibnamefont {Ueda}}, \ and\ \bibinfo {author} {\bibfnamefont
  {Y.}~\bibnamefont {Katsumura}},\ }\href@noop {} {\bibfield  {journal}
  {\bibinfo  {journal} {Rad. Phys. and Chem.}\ }\textbf {\bibinfo {volume}
  {77}},\ \bibinfo {pages} {1176} (\bibinfo {year} {2008})}\BibitemShut
  {NoStop}%
\bibitem [{\citenamefont {He}\ \emph {et~al.}(2013)\citenamefont {He},
  \citenamefont {Hou}, \citenamefont {Easter}, \citenamefont {Faure},
  \citenamefont {Krushelnick}, \citenamefont {Nees},\ and\ \citenamefont
  {Thomas}}]{he13}%
  \BibitemOpen
  \bibfield  {author} {\bibinfo {author} {\bibfnamefont {Z.-H.}\ \bibnamefont
  {He}}, \bibinfo {author} {\bibfnamefont {B.}~\bibnamefont {Hou}}, \bibinfo
  {author} {\bibfnamefont {J.~H.}\ \bibnamefont {Easter}}, \bibinfo {author}
  {\bibfnamefont {J.}~\bibnamefont {Faure}}, \bibinfo {author} {\bibfnamefont
  {K.}~\bibnamefont {Krushelnick}}, \bibinfo {author} {\bibfnamefont {J.~A.}\
  \bibnamefont {Nees}}, \ and\ \bibinfo {author} {\bibfnamefont {A.~G.~R.}\
  \bibnamefont {Thomas}},\ }\href@noop {} {\bibfield  {journal} {\bibinfo
  {journal} {New J. Phys.}\ }\textbf {\bibinfo {volume} {15}},\ \bibinfo
  {pages} {053016} (\bibinfo {year} {2013})}\BibitemShut {NoStop}%
\bibitem [{\citenamefont {Beaurepaire}\ \emph {et~al.}(2015)\citenamefont
  {Beaurepaire}, \citenamefont {Vernier}, \citenamefont {Bocoum}, \citenamefont
  {B\"ohle}, \citenamefont {Jullien}, \citenamefont {Rousseau}, \citenamefont
  {Lefrou}, \citenamefont {Douillet}, \citenamefont {Iaquaniello},
  \citenamefont {Lopez-Martens}, \citenamefont {Lifschitz},\ and\ \citenamefont
  {Faure}}]{beau15}%
  \BibitemOpen
  \bibfield  {author} {\bibinfo {author} {\bibfnamefont {B.}~\bibnamefont
  {Beaurepaire}}, \bibinfo {author} {\bibfnamefont {A.}~\bibnamefont
  {Vernier}}, \bibinfo {author} {\bibfnamefont {M.}~\bibnamefont {Bocoum}},
  \bibinfo {author} {\bibfnamefont {F.}~\bibnamefont {B\"ohle}}, \bibinfo
  {author} {\bibfnamefont {A.}~\bibnamefont {Jullien}}, \bibinfo {author}
  {\bibfnamefont {J.-P.}\ \bibnamefont {Rousseau}}, \bibinfo {author}
  {\bibfnamefont {T.}~\bibnamefont {Lefrou}}, \bibinfo {author} {\bibfnamefont
  {D.}~\bibnamefont {Douillet}}, \bibinfo {author} {\bibfnamefont
  {G.}~\bibnamefont {Iaquaniello}}, \bibinfo {author} {\bibfnamefont
  {R.}~\bibnamefont {Lopez-Martens}}, \bibinfo {author} {\bibfnamefont
  {A.}~\bibnamefont {Lifschitz}}, \ and\ \bibinfo {author} {\bibfnamefont
  {J.}~\bibnamefont {Faure}},\ }\href@noop {} {\bibfield  {journal} {\bibinfo
  {journal} {Phys. Rev. X}\ }\textbf {\bibinfo {volume} {5}},\ \bibinfo {pages}
  {031012} (\bibinfo {year} {2015})}\BibitemShut {NoStop}%
\bibitem [{\citenamefont {Salehi}\ \emph {et~al.}(2017)\citenamefont {Salehi},
  \citenamefont {Goers}, \citenamefont {Hine}, \citenamefont {Feder},
  \citenamefont {Kuk}, \citenamefont {Miao}, \citenamefont {Woodbury},
  \citenamefont {Kim},\ and\ \citenamefont {Milchberg}}]{salehi17}%
  \BibitemOpen
  \bibfield  {author} {\bibinfo {author} {\bibfnamefont {F.}~\bibnamefont
  {Salehi}}, \bibinfo {author} {\bibfnamefont {A.~J.}\ \bibnamefont {Goers}},
  \bibinfo {author} {\bibfnamefont {G.~A.}\ \bibnamefont {Hine}}, \bibinfo
  {author} {\bibfnamefont {L.}~\bibnamefont {Feder}}, \bibinfo {author}
  {\bibfnamefont {D.}~\bibnamefont {Kuk}}, \bibinfo {author} {\bibfnamefont
  {B.}~\bibnamefont {Miao}}, \bibinfo {author} {\bibfnamefont {D.}~\bibnamefont
  {Woodbury}}, \bibinfo {author} {\bibfnamefont {K.~Y.}\ \bibnamefont {Kim}}, \
  and\ \bibinfo {author} {\bibfnamefont {H.~M.}\ \bibnamefont {Milchberg}},\
  }\href@noop {} {\bibfield  {journal} {\bibinfo  {journal} {Opt. Lett.}\
  }\textbf {\bibinfo {volume} {42}},\ \bibinfo {pages} {215} (\bibinfo {year}
  {2017})}\BibitemShut {NoStop}%
\bibitem [{\citenamefont {Pukhov}\ and\ \citenamefont
  {{Meyer-ter-Vehn}}(2002)}]{pukh02}%
  \BibitemOpen
  \bibfield  {author} {\bibinfo {author} {\bibfnamefont {A.}~\bibnamefont
  {Pukhov}}\ and\ \bibinfo {author} {\bibfnamefont {J.}~\bibnamefont
  {{Meyer-ter-Vehn}}},\ }\href@noop {} {\bibfield  {journal} {\bibinfo
  {journal} {Appl. Phys. B}\ }\textbf {\bibinfo {volume} {74}},\ \bibinfo
  {pages} {355} (\bibinfo {year} {2002})}\BibitemShut {NoStop}%
\bibitem [{\citenamefont {Gu\'enot}\ \emph {et~al.}(2017)\citenamefont
  {Gu\'enot}, \citenamefont {Gustas}, \citenamefont {Vernier}, \citenamefont
  {Beaurepaire}, \citenamefont {B{\"o}hle}, \citenamefont {Bocoum},
  \citenamefont {Lozano}, \citenamefont {Jullien}, \citenamefont
  {Lopez-Martens}, \citenamefont {Lifschitz},\ and\ \citenamefont
  {Faure}}]{guenot17}%
  \BibitemOpen
  \bibfield  {author} {\bibinfo {author} {\bibfnamefont {D.}~\bibnamefont
  {Gu\'enot}}, \bibinfo {author} {\bibfnamefont {D.}~\bibnamefont {Gustas}},
  \bibinfo {author} {\bibfnamefont {A.}~\bibnamefont {Vernier}}, \bibinfo
  {author} {\bibfnamefont {B.}~\bibnamefont {Beaurepaire}}, \bibinfo {author}
  {\bibfnamefont {F.}~\bibnamefont {B{\"o}hle}}, \bibinfo {author}
  {\bibfnamefont {M.}~\bibnamefont {Bocoum}}, \bibinfo {author} {\bibfnamefont
  {M.}~\bibnamefont {Lozano}}, \bibinfo {author} {\bibfnamefont
  {A.}~\bibnamefont {Jullien}}, \bibinfo {author} {\bibfnamefont
  {R.}~\bibnamefont {Lopez-Martens}}, \bibinfo {author} {\bibfnamefont
  {A.}~\bibnamefont {Lifschitz}}, \ and\ \bibinfo {author} {\bibfnamefont
  {J.}~\bibnamefont {Faure}},\ }\href@noop {} {\bibfield  {journal} {\bibinfo
  {journal} {Nat. Photon.}\ }\textbf {\bibinfo {volume} {11}},\ \bibinfo
  {pages} {293} (\bibinfo {year} {2017})}\BibitemShut {NoStop}%
\bibitem [{\citenamefont {Pak}\ \emph {et~al.}(2010)\citenamefont {Pak},
  \citenamefont {Marsh}, \citenamefont {Martins}, \citenamefont {Lu},
  \citenamefont {Mori},\ and\ \citenamefont {Joshi}}]{pak10}%
  \BibitemOpen
  \bibfield  {author} {\bibinfo {author} {\bibfnamefont {A.}~\bibnamefont
  {Pak}}, \bibinfo {author} {\bibfnamefont {K.~A.}\ \bibnamefont {Marsh}},
  \bibinfo {author} {\bibfnamefont {S.~F.}\ \bibnamefont {Martins}}, \bibinfo
  {author} {\bibfnamefont {W.}~\bibnamefont {Lu}}, \bibinfo {author}
  {\bibfnamefont {W.~B.}\ \bibnamefont {Mori}}, \ and\ \bibinfo {author}
  {\bibfnamefont {C.}~\bibnamefont {Joshi}},\ }\href {\doibase
  10.1103/PhysRevLett.104.025003} {\bibfield  {journal} {\bibinfo  {journal}
  {Phys. Rev. Lett.}\ }\textbf {\bibinfo {volume} {104}},\ \bibinfo {pages}
  {025003} (\bibinfo {year} {2010})}\BibitemShut {NoStop}%
\bibitem [{\citenamefont {McGuffey}\ \emph {et~al.}(2010)\citenamefont
  {McGuffey}, \citenamefont {Thomas}, \citenamefont {Schumaker}, \citenamefont
  {Matsuoka}, \citenamefont {Chvykov}, \citenamefont {Dollar}, \citenamefont
  {Kalintchenko}, \citenamefont {Yanovsky}, \citenamefont {Maksimchuk},
  \citenamefont {Krushelnick}, \citenamefont {Bychenkov}, \citenamefont
  {Glazyrin},\ and\ \citenamefont {Karpeev}}]{mcgu10}%
  \BibitemOpen
  \bibfield  {author} {\bibinfo {author} {\bibfnamefont {C.}~\bibnamefont
  {McGuffey}}, \bibinfo {author} {\bibfnamefont {A.~G.~R.}\ \bibnamefont
  {Thomas}}, \bibinfo {author} {\bibfnamefont {W.}~\bibnamefont {Schumaker}},
  \bibinfo {author} {\bibfnamefont {T.}~\bibnamefont {Matsuoka}}, \bibinfo
  {author} {\bibfnamefont {V.}~\bibnamefont {Chvykov}}, \bibinfo {author}
  {\bibfnamefont {F.~J.}\ \bibnamefont {Dollar}}, \bibinfo {author}
  {\bibfnamefont {G.}~\bibnamefont {Kalintchenko}}, \bibinfo {author}
  {\bibfnamefont {V.}~\bibnamefont {Yanovsky}}, \bibinfo {author}
  {\bibfnamefont {A.}~\bibnamefont {Maksimchuk}}, \bibinfo {author}
  {\bibfnamefont {K.}~\bibnamefont {Krushelnick}}, \bibinfo {author}
  {\bibfnamefont {V.~Y.}\ \bibnamefont {Bychenkov}}, \bibinfo {author}
  {\bibfnamefont {I.~V.}\ \bibnamefont {Glazyrin}}, \ and\ \bibinfo {author}
  {\bibfnamefont {A.~V.}\ \bibnamefont {Karpeev}},\ }\href {\doibase
  10.1103/PhysRevLett.104.025004} {\bibfield  {journal} {\bibinfo  {journal}
  {Phys. Rev. Lett.}\ }\textbf {\bibinfo {volume} {104}},\ \bibinfo {pages}
  {025004} (\bibinfo {year} {2010})}\BibitemShut {NoStop}%
\bibitem [{\citenamefont {Lundh}\ \emph {et~al.}(2013)\citenamefont {Lundh},
  \citenamefont {Rechatin}, \citenamefont {Lim}, \citenamefont {Malka},\ and\
  \citenamefont {Faure}}]{lund13}%
  \BibitemOpen
  \bibfield  {author} {\bibinfo {author} {\bibfnamefont {O.}~\bibnamefont
  {Lundh}}, \bibinfo {author} {\bibfnamefont {C.}~\bibnamefont {Rechatin}},
  \bibinfo {author} {\bibfnamefont {J.}~\bibnamefont {Lim}}, \bibinfo {author}
  {\bibfnamefont {V.}~\bibnamefont {Malka}}, \ and\ \bibinfo {author}
  {\bibfnamefont {J.}~\bibnamefont {Faure}},\ }\href@noop {} {\bibfield
  {journal} {\bibinfo  {journal} {Phys. Rev. Lett.}\ }\textbf {\bibinfo
  {volume} {110}},\ \bibinfo {pages} {219} (\bibinfo {year}
  {2013})}\BibitemShut {NoStop}%
\bibitem [{\citenamefont {Mangles}\ \emph {et~al.}(2012)\citenamefont
  {Mangles}, \citenamefont {Genoud}, \citenamefont {Bloom}, \citenamefont
  {Burza}, \citenamefont {Najmudin}, \citenamefont {Persson}, \citenamefont
  {Svensson}, \citenamefont {Thomas},\ and\ \citenamefont
  {Wahlstr{\"o}m}}]{mang12}%
  \BibitemOpen
  \bibfield  {author} {\bibinfo {author} {\bibfnamefont {S.~P.~D.}\
  \bibnamefont {Mangles}}, \bibinfo {author} {\bibfnamefont {G.}~\bibnamefont
  {Genoud}}, \bibinfo {author} {\bibfnamefont {M.~S.}\ \bibnamefont {Bloom}},
  \bibinfo {author} {\bibfnamefont {M.}~\bibnamefont {Burza}}, \bibinfo
  {author} {\bibfnamefont {Z.}~\bibnamefont {Najmudin}}, \bibinfo {author}
  {\bibfnamefont {A.}~\bibnamefont {Persson}}, \bibinfo {author} {\bibfnamefont
  {K.}~\bibnamefont {Svensson}}, \bibinfo {author} {\bibfnamefont {A.~G.~R.}\
  \bibnamefont {Thomas}}, \ and\ \bibinfo {author} {\bibfnamefont {C.-G.}\
  \bibnamefont {Wahlstr{\"o}m}},\ }\href@noop {} {\bibfield  {journal}
  {\bibinfo  {journal} {Phys. Rev. ST Accel. Beams}\ }\textbf {\bibinfo
  {volume} {15}},\ \bibinfo {pages} {011302} (\bibinfo {year}
  {2012})}\BibitemShut {NoStop}%
\bibitem [{\citenamefont {Mangles}\ \emph {et~al.}(2004)\citenamefont
  {Mangles}, \citenamefont {Murphy}, \citenamefont {Najmudin}, \citenamefont
  {Thomas}, \citenamefont {Collier}, \citenamefont {Dangor}, \citenamefont
  {Divall}, \citenamefont {Foster}, \citenamefont {Gallacher}, \citenamefont
  {Hooker}, \citenamefont {Jaroszynski}, \citenamefont {Langley}, \citenamefont
  {Mori}, \citenamefont {Norreys}, \citenamefont {Tsung}, \citenamefont
  {Walton},\ and\ \citenamefont {Krushelnick}}]{mang04}%
  \BibitemOpen
  \bibfield  {author} {\bibinfo {author} {\bibfnamefont {S.~P.~D.}\
  \bibnamefont {Mangles}}, \bibinfo {author} {\bibfnamefont {C.~D.}\
  \bibnamefont {Murphy}}, \bibinfo {author} {\bibfnamefont {Z.}~\bibnamefont
  {Najmudin}}, \bibinfo {author} {\bibfnamefont {A.~G.~R.}\ \bibnamefont
  {Thomas}}, \bibinfo {author} {\bibfnamefont {J.~L.}\ \bibnamefont {Collier}},
  \bibinfo {author} {\bibfnamefont {A.~E.}\ \bibnamefont {Dangor}}, \bibinfo
  {author} {\bibfnamefont {E.~J.}\ \bibnamefont {Divall}}, \bibinfo {author}
  {\bibfnamefont {P.~S.}\ \bibnamefont {Foster}}, \bibinfo {author}
  {\bibfnamefont {J.~G.}\ \bibnamefont {Gallacher}}, \bibinfo {author}
  {\bibfnamefont {C.~J.}\ \bibnamefont {Hooker}}, \bibinfo {author}
  {\bibfnamefont {D.~A.}\ \bibnamefont {Jaroszynski}}, \bibinfo {author}
  {\bibfnamefont {A.~J.}\ \bibnamefont {Langley}}, \bibinfo {author}
  {\bibfnamefont {W.~B.}\ \bibnamefont {Mori}}, \bibinfo {author}
  {\bibfnamefont {P.~A.}\ \bibnamefont {Norreys}}, \bibinfo {author}
  {\bibfnamefont {F.~S.}\ \bibnamefont {Tsung}}, \bibinfo {author}
  {\bibfnamefont {B.~R.}\ \bibnamefont {Walton}}, \ and\ \bibinfo {author}
  {\bibfnamefont {K.}~\bibnamefont {Krushelnick}},\ }\href@noop {} {\bibfield
  {journal} {\bibinfo  {journal} {Nature}\ }\textbf {\bibinfo {volume} {431}},\
  \bibinfo {pages} {535} (\bibinfo {year} {2004})}\BibitemShut {NoStop}%
\bibitem [{\citenamefont {Geddes}\ \emph {et~al.}(2004)\citenamefont {Geddes},
  \citenamefont {T\'{o}th}, \citenamefont {{van Tilborg}}, \citenamefont
  {Esarey}, \citenamefont {Schroeder}, \citenamefont {Bruhwiler}, \citenamefont
  {Nieter}, \citenamefont {Cary},\ and\ \citenamefont {Leemans}}]{gedd04}%
  \BibitemOpen
  \bibfield  {author} {\bibinfo {author} {\bibfnamefont {C.~G.~R.}\
  \bibnamefont {Geddes}}, \bibinfo {author} {\bibfnamefont {C.}~\bibnamefont
  {T\'{o}th}}, \bibinfo {author} {\bibfnamefont {J.}~\bibnamefont {{van
  Tilborg}}}, \bibinfo {author} {\bibfnamefont {E.}~\bibnamefont {Esarey}},
  \bibinfo {author} {\bibfnamefont {C.~B.}\ \bibnamefont {Schroeder}}, \bibinfo
  {author} {\bibfnamefont {D.}~\bibnamefont {Bruhwiler}}, \bibinfo {author}
  {\bibfnamefont {C.}~\bibnamefont {Nieter}}, \bibinfo {author} {\bibfnamefont
  {J.}~\bibnamefont {Cary}}, \ and\ \bibinfo {author} {\bibfnamefont {W.~P.}\
  \bibnamefont {Leemans}},\ }\href@noop {} {\bibfield  {journal} {\bibinfo
  {journal} {Nature}\ }\textbf {\bibinfo {volume} {431}},\ \bibinfo {pages}
  {538} (\bibinfo {year} {2004})}\BibitemShut {NoStop}%
\bibitem [{\citenamefont {Faure}\ \emph {et~al.}(2004)\citenamefont {Faure},
  \citenamefont {Glinec}, \citenamefont {Pukhov}, \citenamefont {Kiselev},
  \citenamefont {Gordienko}, \citenamefont {Lefebvre}, \citenamefont
  {Rousseau}, \citenamefont {Burgy},\ and\ \citenamefont {Malka}}]{faur04}%
  \BibitemOpen
  \bibfield  {author} {\bibinfo {author} {\bibfnamefont {J.}~\bibnamefont
  {Faure}}, \bibinfo {author} {\bibfnamefont {Y.}~\bibnamefont {Glinec}},
  \bibinfo {author} {\bibfnamefont {A.}~\bibnamefont {Pukhov}}, \bibinfo
  {author} {\bibfnamefont {S.}~\bibnamefont {Kiselev}}, \bibinfo {author}
  {\bibfnamefont {S.}~\bibnamefont {Gordienko}}, \bibinfo {author}
  {\bibfnamefont {E.}~\bibnamefont {Lefebvre}}, \bibinfo {author}
  {\bibfnamefont {J.-P.}\ \bibnamefont {Rousseau}}, \bibinfo {author}
  {\bibfnamefont {F.}~\bibnamefont {Burgy}}, \ and\ \bibinfo {author}
  {\bibfnamefont {V.}~\bibnamefont {Malka}},\ }\href@noop {} {\bibfield
  {journal} {\bibinfo  {journal} {Nature}\ }\textbf {\bibinfo {volume} {431}},\
  \bibinfo {pages} {541} (\bibinfo {year} {2004})}\BibitemShut {NoStop}%
\bibitem [{\citenamefont {Lu}\ \emph {et~al.}(2007)\citenamefont {Lu},
  \citenamefont {Tzoufras}, \citenamefont {Joshi}, \citenamefont {Tsung},
  \citenamefont {Mori}, \citenamefont {Vieira}, \citenamefont {Fonseca},\ and\
  \citenamefont {Silva}}]{lu07}%
  \BibitemOpen
  \bibfield  {author} {\bibinfo {author} {\bibfnamefont {W.}~\bibnamefont
  {Lu}}, \bibinfo {author} {\bibfnamefont {M.}~\bibnamefont {Tzoufras}},
  \bibinfo {author} {\bibfnamefont {C.}~\bibnamefont {Joshi}}, \bibinfo
  {author} {\bibfnamefont {F.~S.}\ \bibnamefont {Tsung}}, \bibinfo {author}
  {\bibfnamefont {W.~B.}\ \bibnamefont {Mori}}, \bibinfo {author}
  {\bibfnamefont {J.}~\bibnamefont {Vieira}}, \bibinfo {author} {\bibfnamefont
  {R.~A.}\ \bibnamefont {Fonseca}}, \ and\ \bibinfo {author} {\bibfnamefont
  {L.~O.}\ \bibnamefont {Silva}},\ }\href@noop {} {\bibfield  {journal}
  {\bibinfo  {journal} {Phys. Rev. ST Accel. Beams}\ }\textbf {\bibinfo
  {volume} {10}},\ \bibinfo {pages} {061301} (\bibinfo {year}
  {2007})}\BibitemShut {NoStop}%
\bibitem [{\citenamefont {Beaurepaire}\ \emph {et~al.}(2014)\citenamefont
  {Beaurepaire}, \citenamefont {Lifschitz},\ and\ \citenamefont
  {Faure}}]{beau14}%
  \BibitemOpen
  \bibfield  {author} {\bibinfo {author} {\bibfnamefont {B.}~\bibnamefont
  {Beaurepaire}}, \bibinfo {author} {\bibfnamefont {A.}~\bibnamefont
  {Lifschitz}}, \ and\ \bibinfo {author} {\bibfnamefont {J.}~\bibnamefont
  {Faure}},\ }\href@noop {} {\bibfield  {journal} {\bibinfo  {journal} {New J.
  Phys.}\ }\textbf {\bibinfo {volume} {16}},\ \bibinfo {pages} {023023}
  (\bibinfo {year} {2014})}\BibitemShut {NoStop}%
\bibitem [{\citenamefont {Rae}(1993)}]{rae93}%
  \BibitemOpen
  \bibfield  {author} {\bibinfo {author} {\bibfnamefont {S.~C.}\ \bibnamefont
  {Rae}},\ }\href@noop {} {\bibfield  {journal} {\bibinfo  {journal} {Opt.
  Comm.}\ }\textbf {\bibinfo {volume} {97}},\ \bibinfo {pages} {25} (\bibinfo
  {year} {1993})}\BibitemShut {NoStop}%
\bibitem [{\citenamefont {Schmid}\ and\ \citenamefont {Veisz}(2012)}]{schm12}%
  \BibitemOpen
  \bibfield  {author} {\bibinfo {author} {\bibfnamefont {K.}~\bibnamefont
  {Schmid}}\ and\ \bibinfo {author} {\bibfnamefont {L.}~\bibnamefont {Veisz}},\
  }\href@noop {} {\bibfield  {journal} {\bibinfo  {journal} {Rev. Sci.
  Intstrum.}\ }\textbf {\bibinfo {volume} {84}},\ \bibinfo {pages} {053304}
  (\bibinfo {year} {2012})}\BibitemShut {NoStop}%
\bibitem [{\citenamefont {B\"{o}hle}\ \emph {et~al.}(2014)\citenamefont
  {B\"{o}hle}, \citenamefont {Kretschmar}, \citenamefont {Jullien},
  \citenamefont {Kovacs}, \citenamefont {Miranda}, \citenamefont {Romero},
  \citenamefont {Crespo}, \citenamefont {Morgner}, \citenamefont {Simon},
  \citenamefont {Lopez-Martens},\ and\ \citenamefont {Nagy}}]{boeh14}%
  \BibitemOpen
  \bibfield  {author} {\bibinfo {author} {\bibfnamefont {F.}~\bibnamefont
  {B\"{o}hle}}, \bibinfo {author} {\bibfnamefont {M.}~\bibnamefont
  {Kretschmar}}, \bibinfo {author} {\bibfnamefont {A.}~\bibnamefont {Jullien}},
  \bibinfo {author} {\bibfnamefont {M.}~\bibnamefont {Kovacs}}, \bibinfo
  {author} {\bibfnamefont {M.}~\bibnamefont {Miranda}}, \bibinfo {author}
  {\bibfnamefont {R.}~\bibnamefont {Romero}}, \bibinfo {author} {\bibfnamefont
  {H.}~\bibnamefont {Crespo}}, \bibinfo {author} {\bibfnamefont
  {U.}~\bibnamefont {Morgner}}, \bibinfo {author} {\bibfnamefont
  {P.}~\bibnamefont {Simon}}, \bibinfo {author} {\bibfnamefont
  {R.}~\bibnamefont {Lopez-Martens}}, \ and\ \bibinfo {author} {\bibfnamefont
  {T.}~\bibnamefont {Nagy}},\ }\href@noop {} {\bibfield  {journal} {\bibinfo
  {journal} {Laser Phys. Lett.}\ }\textbf {\bibinfo {volume} {11}},\ \bibinfo
  {pages} {095401} (\bibinfo {year} {2014})}\BibitemShut {NoStop}%
\bibitem [{\citenamefont {Lifschitz}\ \emph {et~al.}(2009)\citenamefont
  {Lifschitz}, \citenamefont {Davoine}, \citenamefont {Lefebvre}, \citenamefont
  {Faure}, \citenamefont {Rechatin},\ and\ \citenamefont {Malka}}]{lifs09}%
  \BibitemOpen
  \bibfield  {author} {\bibinfo {author} {\bibfnamefont {A.}~\bibnamefont
  {Lifschitz}}, \bibinfo {author} {\bibfnamefont {X.}~\bibnamefont {Davoine}},
  \bibinfo {author} {\bibfnamefont {E.}~\bibnamefont {Lefebvre}}, \bibinfo
  {author} {\bibfnamefont {J.}~\bibnamefont {Faure}}, \bibinfo {author}
  {\bibfnamefont {C.}~\bibnamefont {Rechatin}}, \ and\ \bibinfo {author}
  {\bibfnamefont {V.}~\bibnamefont {Malka}},\ }\href@noop {} {\bibfield
  {journal} {\bibinfo  {journal} {J. Comp. Phys.}\ }\textbf {\bibinfo {volume}
  {228}},\ \bibinfo {pages} {1803} (\bibinfo {year} {2009})}\BibitemShut
  {NoStop}%
\bibitem [{\citenamefont {Glinec}\ \emph {et~al.}(2005)\citenamefont {Glinec},
  \citenamefont {Faure}, \citenamefont {Dain}, \citenamefont {Darbon},
  \citenamefont {Hosokai}, \citenamefont {Santos}, \citenamefont {Lefebvre},
  \citenamefont {Rousseau}, \citenamefont {Burgy}, \citenamefont {Mercier},\
  and\ \citenamefont {Malka}}]{glin05}%
  \BibitemOpen
  \bibfield  {author} {\bibinfo {author} {\bibfnamefont {Y.}~\bibnamefont
  {Glinec}}, \bibinfo {author} {\bibfnamefont {J.}~\bibnamefont {Faure}},
  \bibinfo {author} {\bibfnamefont {L.~L.}\ \bibnamefont {Dain}}, \bibinfo
  {author} {\bibfnamefont {S.}~\bibnamefont {Darbon}}, \bibinfo {author}
  {\bibfnamefont {T.}~\bibnamefont {Hosokai}}, \bibinfo {author} {\bibfnamefont
  {J.~J.}\ \bibnamefont {Santos}}, \bibinfo {author} {\bibfnamefont
  {E.}~\bibnamefont {Lefebvre}}, \bibinfo {author} {\bibfnamefont {J.~P.}\
  \bibnamefont {Rousseau}}, \bibinfo {author} {\bibfnamefont {F.}~\bibnamefont
  {Burgy}}, \bibinfo {author} {\bibfnamefont {B.}~\bibnamefont {Mercier}}, \
  and\ \bibinfo {author} {\bibfnamefont {V.}~\bibnamefont {Malka}},\ }\href
  {\doibase 10.1103/PhysRevLett.94.025003} {\bibfield  {journal} {\bibinfo
  {journal} {Phys. Rev. Lett.}\ }\textbf {\bibinfo {volume} {94}},\ \bibinfo
  {eid} {025003} (\bibinfo {year} {2005})}\BibitemShut {NoStop}%
\bibitem [{\citenamefont {Faure}\ \emph {et~al.}(2016)\citenamefont {Faure},
  \citenamefont {van~der Geer}, \citenamefont {Beaurepaire}, \citenamefont
  {Gall\'e}, \citenamefont {Vernier},\ and\ \citenamefont
  {Lifschitz}}]{faure16}%
  \BibitemOpen
  \bibfield  {author} {\bibinfo {author} {\bibfnamefont {J.}~\bibnamefont
  {Faure}}, \bibinfo {author} {\bibfnamefont {B.}~\bibnamefont {van~der Geer}},
  \bibinfo {author} {\bibfnamefont {B.}~\bibnamefont {Beaurepaire}}, \bibinfo
  {author} {\bibfnamefont {G.}~\bibnamefont {Gall\'e}}, \bibinfo {author}
  {\bibfnamefont {A.}~\bibnamefont {Vernier}}, \ and\ \bibinfo {author}
  {\bibfnamefont {A.}~\bibnamefont {Lifschitz}},\ }\href@noop {} {\bibfield
  {journal} {\bibinfo  {journal} {Phys. Rev. Accel. Beams}\ }\textbf {\bibinfo
  {volume} {19}},\ \bibinfo {pages} {021302} (\bibinfo {year}
  {2016})}\BibitemShut {NoStop}%
\end{thebibliography}

%

\end{document}